\documentclass[aps,prx, twocolumn,amsmath, longbibliography, amssymb,superscriptaddress]{revtex4-1}

\usepackage{bm}
\usepackage[retainorgcmds]{IEEEtrantools}
\usepackage{graphicx}
\usepackage{mathrsfs}
\usepackage{amsmath}
\usepackage{amssymb}
\usepackage{color}
\usepackage{amsfonts}
\usepackage{times,txfonts}
\usepackage{nicefrac}
\usepackage[colorlinks=true,linkcolor=blue,urlcolor=blue,citecolor=blue,pdfusetitle]{hyperref}
\usepackage{physics}

\newcommand{\scnd}{2^{\text{nd}}}
\newcommand{\Var}{\text{var}}
\newcommand{\E}{\mathbb{E}}

\newcommand{\trans}{^{\text{T}}}

\newcommand{\nn}{\bar{n} + {1}/{2}}

\begin{document}

\title{Informational steady-states and conditional entropy production in continuously monitored systems: the case of Gaussian systems}
\date{\today}
\author{Alessio Belenchia}
\affiliation{Institut f\"{u}r Theoretische Physik, Eberhard-Karls-Universit\"{a}t T\"{u}bingen, 72076 T\"{u}bingen, Germany}
\affiliation{Centre for Theoretical Atomic, Molecular, and Optical Physics, School of Mathematics and Physics, Queens University, Belfast BT7 1NN, United Kingdom}
\author{Mauro Paternostro}
\affiliation{Centre for Theoretical Atomic, Molecular, and Optical Physics, School of Mathematics and Physics, Queens University, Belfast BT7 1NN, United Kingdom}
\author{Gabriel T. Landi}
\affiliation{Instituto de F\'isica da Universidade de S\~ao Paulo,  05314-970 S\~ao Paulo, Brazil.}

\begin{abstract}
The act of measuring a system has profound consequences of dynamical and thermodynamic nature. In particular, the degree of irreversibility ensuing from a non-equilibrium process is strongly affected by measurements aimed at acquiring information on the state of a system of interest: the conditional and unconditional entropy production, which quantify the degree of irreversibility of the open system's dynamics, are related to each other by clearly interpreted informational quantities. Building on a recently proposed collisional-model framework [G. T. Landi {\it et al.}, arXiv:2103.06247], we investigate the case of continuous-variable information carriers prepared in Gaussian states and undergoing Gaussian processes. We build up a toolbox that fully characterizes the thermodynamics of continuously measured non-equilibrium Gaussian systems and processes, illustrating how the instruments hereby introduced provide key insight into recent experiments on mesoscopic quantum systems [Phys. Rev. Lett, {\bf 125}, 080601 (2020)].
\end{abstract}

\maketitle 

%

%
%
\section{Introduction}
%
%

The measurement process is at the basis of our ability to acquire information on both classical and quantum systems. In the latter case, however, this process introduces unavoidable disturbances into the dynamics of the system, whenever some information can be extracted~\cite{busch2009no}. These considerations have an important impact not only on the dynamics of quantum systems but also on their thermodynamics.

Advances in (quantum) information thermodynamics~\cite{Parrondo2015} have shown how the acquisition of information, and the feedback enabled by it, impact the basic laws of thermodynamics~\cite{Sagawa2008,Toyabe2010,Ito2013,Sagawa2012,Sagawa2013,Funo2013,Koski2014,Elouard2017a,Cottet2017,Naghiloo2018,Buffoni2018,Mohammady2019a,Strasberg2019a,Belenchia2019,Debiossac2019,Beyer2020,Sone2020,Strasberg2020}. This is particularly relevant for the second law of thermodynamics and its generalization in the form of fluctuation theorems. The second law characterises the irreversibility of dynamical stochastic process through the irreversible entropy production and flux rates entering it. These quantities, in turn, are at the basis of the formulation of non-equilibrium and stochastic thermodynamics~\cite{lebon2008understanding}, characterise the efficiency of thermal machines~\cite{cengel2007thermodynamics}, and determine the response of the systems to thermodynamic forces~\cite{gallavotti2013statistical}.

In laboratory practice, quantum systems are often interrogated in a continuous fashion by way of weak, indirect measurements on ancilliary systems~\cite{Wiseman2009,Jacobs2014,Serafini2017}. The chief example of this is presented by cavity optomechanics~\cite{Aspelmeyer2014} where the properties of a mechanical system, and the cavity field, are inferred by measuring the electromagnetic field leaking out from the cavity mirrors with dyne-type continuous measurements. In view of the ubiquitous character of continuously monitored quantum system, a comprehensive theory describing their thermodynamics would advance our understanding of the interplay between information and dissipation in these open quantum systems.

Several works in recent years have addressed the thermodynamics of measured systems, also in the case of continuous measurements~\cite{Funo2013,Strasberg2019a,Strasberg2019c,Strasberg2020,Alonso2016,Strasberg2019c,Strasberg2020,Naghiloo2020}. In this work, we take a step further in this direction by employing a recently proposed collisional model framework~\cite{firstpaper} for the analysis of the information thermodynamics of Gaussian quantum systems and processes subjected to Gaussian quantum measurements. Gaussian quantum systems and measurements are ubiquitous nowadays in quantum laboratories where quadratic Hamiltonians are easily implemented and accurately describe  optical, atomic, and mechanical systems -- and their interaction -- in a variety of technologically relevant situations~\cite{Ferraro2005,Wiseman2009,PirlaRMP,Genoni2016,Serafini2017}. Furthermore, Gaussian quantum measurements, like homodyne and heterodyne quadrature detection, are commonly implemented and allow for exquisite quantum control of these systems. 

While employing the framework of~\cite{firstpaper}, this work is also a natural progression of~\cite{Belenchia2019}, where we put forth a semiclassical, phase-space based thermodynamic framework for continuously measured Gaussian systems. The same framework~\cite{Belenchia2019}, has also been recently used in~\cite{Rossi2020} to experimentally assess the conditional $2^\text{nd}$ law in an optomechanical system. In contrast to~\cite{Belenchia2019}, which was based on quantum phase-space methods, here we develop a microscopic collisional description of the thermodynamics of the same systems. This is motivated by the fact that the results in~\cite{Belenchia2019}, in addition to being semiclassical, are formulated solely in terms of the stochastic master equation obeyed by the system. As such, they do not demand an explicit model of the environment, but only the knowledge of the open dynamics it produces. While this last aspect could seem advantageous, there has been increasing evidence that a proper formulation of thermodynamics in the quantum regime is only possible if information on the environment and its interaction with the system is provided~\cite{Landi2020a}. Indeed, reduced descriptions based only on the master equation can manifest apparent violations of the $\scnd$ law~\cite{Levy2014}, something which can only be resolved by introducing a specific model of the environment~\cite{DeChiara2018}. Our efforts in modeling the experiments in~\cite{Rossi2020} also strongly corroborate this view. In fact, an explicit illustration will be provided in the following, where we will show that models that yield the same type of master equation can have completely different thermodynamics features.

 \begin{figure}
\centering
\includegraphics[width=\columnwidth]{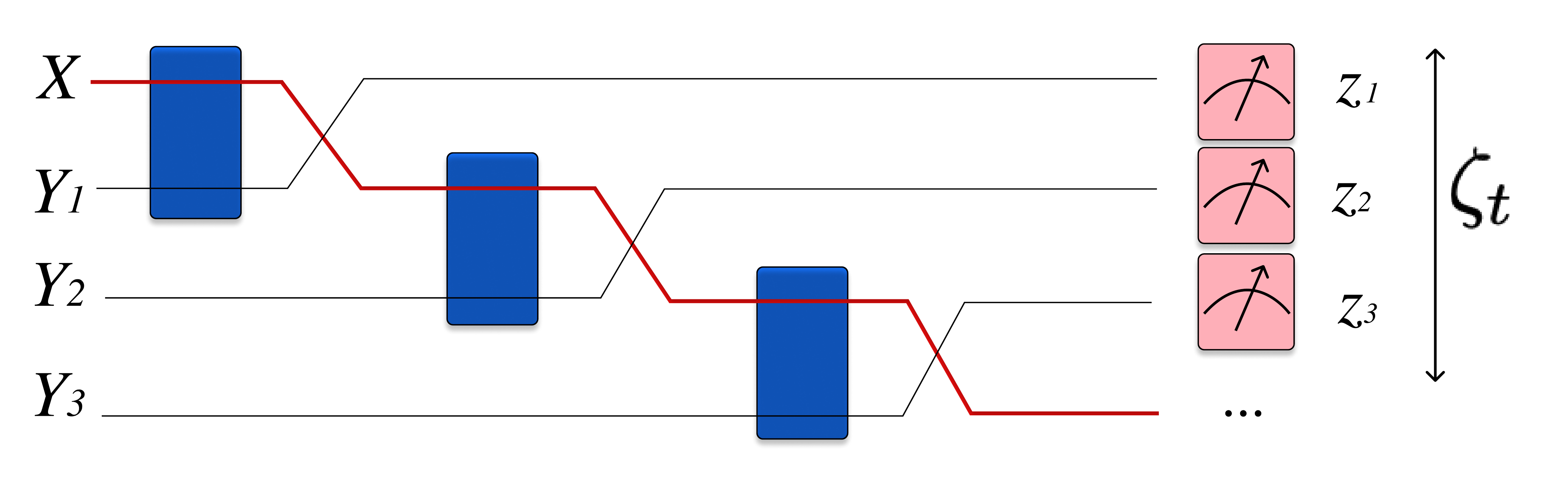}
\caption{\label{fig:drawings}
Circuital representation of the collisional model dynamics for monitored systems. The system ($X$) interacts repeatedly with identically prepared, independent ancillas ($Y_t$) according to the map in Eq.~\eqref{global_map}. The monitoring is introduced through the measurement of the ancillary systems after each collision in accordance with generalized measurement operators $\{M_z\}$ producing a classical (and random) outcome $z_t$.}
\end{figure}

Our paper is organized as follows. 
In Sec.~\ref{sec:setup}, the general collisional model framework for the thermodynamics of continuously monitored systems developed in~\cite{firstpaper} is briefly reviewed establishing the basic formalism. Sec.~\ref{sec:gaussian} is dedicated to the discussion of Gaussian processes and measurement in the collisional model framework. Sec.~\ref{sec:example}, collects some significant example showcasing the application of the formalism and further uses the detailed assessment of the experiment reported in Ref.~\cite{Rossi2020} as a physically motivated case study. Finally, in Sec.~\ref{sec:conc} we draw our conclusions and highlight the perspectives opened by our approach.

%
%
\section{\label{sec:setup}Information and thermodynamics of continuously measured collisional models}
%
%

In this Section we discuss the elements underpinning the collisional model of continuously monitored systems ($\text{CM}^2$) construction and the analysis of its informational and thermodynamic features. This will help us setting the context of the study -- dedicated to Gaussian bosonic systems -- reported in Secs.~\ref{sec:gaussian} and \ref{sec:example}. An extended discussion on the topic of this Section is in Ref.~\cite{firstpaper} to which we refer for further details.  

\subsection{Construction of $\text{CM}^2$}
We consider a system $X$ with initial density matrix $\rho_{X_0}$ interacting sequentially with independent and identically prepared (iid) ancillae $Y_1$, $Y_2\ldots$, all prepared always in the same state $\rho_{Y_t} = \rho_Y$. We discretize time and label the corresponding units as $t = 0,1,2,3,\ldots$. A unitary $U_{t}$ acting on $X$ and $Y_{t}$ (a {\it collision}) takes the system from time ${t-1}$ to ${t}$ as
\begin{equation}\label{global_map}
\rho_{X_{t}Y_{t}'} = U_t(\rho_{X_{t-1}} \otimes \rho_{Y_t}) U_t^\dagger, 
\end{equation}
where $Y_t'$ refers to the state of ancilla $Y_t$ after the event.
Information on the state of the system is acquired by measuring the  $\rho_{Y_t}'$, the state of of the ancilla after the collision.
Such measurement is described by a set of generalized measurement operators $\{M_z\}$, satisfying $\sum_z M_z^\dagger M_z = \openone$ and the associated probability for outcome $z_t$ is 
\begin{equation}\label{Pz}
P(z_t) = \tr\Big\{ M_{z_t} \rho_{Y_t}'  M_{z_t}^\dagger \Big\}.
\end{equation}
After its respective collision, ancilla $Y_t'$ will not participate again in the dynamics, while a fresh ancilla $Y_{t+1}$ is introduced and the sequence repeated. 
This allows us to build a set of time-ordered measurement records (see Fig.~\ref{fig:drawings})
\begin{equation}\label{zeta}
\zeta_t = (z_1, \ldots, z_t). 
\end{equation}
In a sense, $\zeta_t$ embodies the ``integrated'' information on $X$, while $z_t$ is a \emph{differential information gain} associated only with the step $X_{t-1} \to X_{t}$. 

The joint distribution $P(\zeta_t)$ is given by~\footnote{Note that, as the measurements act only on those ancillae that no longer participate in the dynamics, it is irrelevant whether the measurement $M_{z_t}$ occurs before or after the next evolution with $Y_{t+1}$. Intuitively, we imagine that the measurement takes place after each collision. But the result would not change at all if we were to measure all ancillae only at the end.}
\begin{equation}
P(\zeta_t) = \tr_{XY_1\ldots Y_{t}} \Big\{ M_{z_{t}} \ldots M_{z_1} \rho_{XY_1\ldots Y_{t}} M_{z_1}^\dagger \ldots M_{z_{t}}^\dagger \Big\}, 
\end{equation}
where 
\begin{equation*}
\rho_{XY_1\ldots Y_{t}} = \left(\Pi^t_{k=1}U_{k}\right)\left( \rho_{X_0} \bigotimes^t_{j=1} \rho_{Y_j} \right)\left( \Pi^t_{k=1}U_{k}\right)^\dagger. 
\end{equation*}

It is now convenient to introduce the outcome-indexed, completely positive, trace non-preserving map
\begin{equation}\label{conditional_map_def}
\mathcal{E}_z(\rho_X) = \tr_Y \Big\{ M_z U (\rho_X \otimes \rho_Y) U^\dagger M_z^\dagger \Big\}, 
\end{equation}
which allows us to define the unnormalized conditional states 
\begin{equation}\label{conditional_map}
\varrho_{X_{t} | \zeta_{t} } = \mathcal{E}_{z_t} \big( \varrho_{X_{t-1}|\zeta_{t-1}}\big)
\end{equation}
with initial condition $\varrho_{X_0|\zeta_0} = \rho_{X_0}$. 
One may readily verify that 
$\tr_X \varrho_{X_t|\zeta_t} = \tr_X \Big\{ \mathcal{E}_{z_{t}} \circ \ldots \circ \mathcal{E}_{z_1} (\rho_{X_0}) \Big\} = P(\zeta_t)$. 
The states $\varrho_{X_t|\zeta_t}$ therefore contain the outcome distribution $P(\zeta_t)$ at any given time.

\subsection{Informational aspects of $\text{CM}^2$}

The mismatch between the information carried by the conditional state of the system at time $t$ and the corresponding unconditional one 
is quantified by the {Holevo information}~\cite{Holevo1973} 
\begin{equation}\label{holevo}
I(X_t \! :\! \zeta_t) := S(X_t) - S(X_t | \zeta_t)= \sum\limits_{\zeta_t} P(\zeta_t) \;D\big( \rho_{X_t|\zeta_t} || \rho_{X_t}\big) \geqslant 0, 
\end{equation}
where $S(X_t)= - \tr \rho_{X_t} \ln \rho_{X_t}$ is the von Neumann entropy of $\rho_{X_t}$, $S(X_t|\zeta_t)= \sum_{\zeta_t} P(\zeta_t) S(\rho_{X_t|\zeta_t})$ is the quantum-classical conditional entropy average over all trajectories set by the sequence of collisions, and $D(\rho||\sigma) = \tr(\rho \ln \rho - \rho \ln \sigma)$ is the quantum relative entropy. 
Eq.~\eqref{holevo} thus provides a measure of the information known about the system given the measurements performed on the ancillae, which can be interpreted as 
the weighted average of the distance between $\rho_{X_t|\zeta_t}$ and $\rho_{X_t}$~\cite{firstpaper}. 

While Eq.~\eqref{holevo} reflects the {integrated} information acquired about the system, up to time $t$, the conditional 
Holevo information 
 
\begin{equation}\label{gain}
G_t := I_c(X_{t}\!:\! z_{t}| \zeta_{t-1}) {=} I(X_{t}\!:\! \zeta_{t}) {-} I(X_{t}\!:\! \zeta_{t-1}) {=}  S(X_t|\zeta_{t-1}) {-} S(X_t | \zeta_t)
\end{equation}
quantifies the {differential} information gain obtained from a single outcome $z$, at each step.


We then define the differential information loss term 
\begin{equation}\label{loss} 
L_t := I(X_{t-1} \! : \! \zeta_{t-1}) -  I(X_t \! : \! \zeta_{t-1}),
\end{equation}
which can be shown to be strictly non-negative~\cite{firstpaper}.
It is then immediate to prove that 
the trade-off between the gain in information and the (non-negative) measurement backaction quantified by the \emph{information rate} $\Delta I_t := I(X_t \! :\! \zeta_t) - I(X_{t-1} \! :\! \zeta_{t-1})$ 
can be split as 
\begin{equation}\label{Delta_I_splitting}
\Delta I_t= G_t - L_t. 
\end{equation}
In the long-time limit, the system may reach a steady-state such that $\Delta I_\infty = 0$. 
But this might be associated with (mutually balancing) non-null information gain and loss rates.  
When this happens, we say the system has relaxed to an informational steady-state (ISS) such that 
\begin{equation}\label{ISS}
\Delta I_\text{ISS} = 0 \quad \text{ but} \quad G_\text{ISS} = L_\text{ISS} \neq 0. 
\end{equation}
In an ISS, information is continuously acquired and balanced by the noise that is introduced by the measurement.
Crucially, the ISS does \emph{not} mean that $\rho_{X_t|\zeta_t}$ is no longer changing.

\subsection{Thermodynamic aspects of $\text{CM}^2$}\label{3c}

The second law of thermodynamics, in the unconditional case, splits the change in entropy into a contribution stemming from the entropy flow $\Delta\Phi_t^u$ from system to ancilla, plus a contribution $\Delta\Sigma_t^u$ representing the entropy that was irreversibly produced in the process. 
In formal terms 
\begin{equation}\label{2nd}
 S(X_t) - S(X_{t-1})= \Delta \Sigma_t^u - \Delta \Phi_t^u. 
\end{equation}
In thermal processes, the entropy flow $\Delta \Phi_t^u$ is typically linked to the heat flow $\dot{Q}_t$ entering the ancillae through Clausius' expression~\cite{Fermi1956} $\Delta\Phi_t^u = \beta \dot{Q}_t$, where $\beta$ is the inverse temperature of the thermal state the ancillae are in.
But this only holds for thermal ancillae, thus restricting the range of applicability of the formalism. 

Instead, we work within the framework of Ref.~\cite{Esposito2010a} (see also \cite{Manzano2017a,Strasberg2016}), which formulates the entropy production rate in information theoretic terms, as
\begin{equation}\label{sigma}
\Delta \Sigma_t^u= \mathcal{I}(X_t\!:\!Y_{t}') + D(Y_t'||Y_{t}) \geqslant 0, 
\end{equation}
where $\mathcal{I}(X_t\!:\!Y_{t}') = S(\rho_{X_{t}}) + S(\rho_{Y_{t}}')- S(\rho_{X_tY_t'})$ is the quantum mutual information between system and ancilla after Eq.~\eqref{global_map} and $D(Y_t'||Y_t) = D(\rho_{Y_t'} || \rho_{Y_t})$ is the relative entropy between the state of the ancilla before and after the collision.
It follows from this that the entropy flux is additive, and depends solely on the degrees of freedom of the ancilla, according to 

\begin{equation}\label{phi_multiple}
\Delta \Phi_t^u = \sum\limits^N_{j=1} \Delta \Phi_{tj}^u =  \sum\limits^N_{j=1} \tr\Big\{(\rho_{Y_{tj}'} - \rho_{Y_{tj}}) \ln \rho_{Y_{tj}}\Big\},
\end{equation}
This allows one to compute the flux associated to each dissipation channel acting on the system.

Eqs.~(\ref{2nd})-(\ref{phi_multiple}) specify the thermodynamics of the unconditional trajectories $\rho_{X_t}$.
Similar relations can be found in the conditional case ($\rho_{X_t|\zeta_t}$), where the relevant entropy is $S(X_t|\zeta_t)$.
As shown in~\cite{firstpaper}, in this case one finds
\begin{equation}\label{2nd_cond}
 S(X_{t}|\zeta_{t}) - S(X_{t-1}|\zeta_{t-1})=\Delta\Sigma_t^c - \Delta \Phi_t^c,
\end{equation}
which is akin to Eq.~(\ref{2nd}),
but with new quantities $\Delta\Sigma_t^c$ and $\Delta \Phi_t^c$ representing the conditional counterparts of the entropy production and flux. 
However, it can be shown that under mild assumptions, $\Delta\Phi_t^c=\Delta\Phi_t^u$; i.e., the flux is independent of whether or not we condition on the measurement outcomes (see Appendix~\ref{app:gaussian_stuff} for further discussion of this point).

Comparing Eqs.~(\ref{2nd}) and~\eqref{2nd_cond}, then allows one to establish the relation between conditional and unconditional entropy production 
\begin{equation}\label{two_sigmas}
\Delta\Sigma_t^c = \Delta \Sigma_t^u- \Delta I_t.
\end{equation}
The act of conditioning on the measurement outcomes hence changes the entropy production by a quantity associated with the change in the Holevo information. 
This  explicitly connects the information rates with thermodynamics.


The integrated entropy productions $\Sigma_t^\alpha = \sum_{\tau=1}^t \Delta \Sigma_\tau^\alpha$ ($\alpha = u,c$) are also readily found to be related according to 
\begin{equation}\label{integrated_sigmas}
\Sigma_t^c = \Sigma_t^u - I(X_t \! : \! \zeta_t). 
\end{equation}
This  shows  that the difference between conditional and unconditional irreversibility, up to time $t$,  is strictly related to the net information $I(X_t\! : \! \zeta_t)\geqslant 0$. It thus follows that
\begin{equation}\label{integrated_sigma_bound_cu}
\Sigma_t^u \geqslant \Sigma_t^c,
\end{equation}
stating that -- as the indirect measurement approach considered here does not result in direct backaction on the system --{the act of conditioning reduces the irreversibility of a process}. 
In fact, one actually has a stronger bound 
$\Sigma_t^u \geqslant  \Sigma_t^c + \sum_{\tau=1}^t G_\tau$, showing that the mismatch is associated with the integrated information gain.

%
%
\section{\label{sec:gaussian}Continuous variable systems}
%
%

Bosonic systems offer an essential platform for the implementation of continuous measurements, a scenario that is frequently found in quantum optical experiments. In this context, the extensively developed toolbox of continuously monitored Gaussian processes~\cite{Wiseman1993,Wiseman1994,Doherty1999,Genoni2016} can be employed to build an insightful and simple formalism. 
Gaussian scenarios also allow for a more direct comparison with classical models, described in terms of Langevin or Fokker-Planck equations~\cite{Santos2017b}. Indeed, similar considerations on the role of information in thermodynamics have been discussed in this classical context in Ref.~\cite{Horowitz2014b}.

\subsection{\label{ssec:GaussianCM2}Gaussian $\text{CM}^2$s}

We begin by  reviewing  the formalism developed in Ref.~\cite{Genoni2016} for describing the unconditional and conditional dynamics.
The system is described by $N_X$ canonically conjugated operators $\hat{R}_X = (q_1,p_1,\ldots,q_{N_X},p_{N_X})$, while each ancilla is modeled by $N_Y$ variables $\hat{R}_Y = (Q_1,P_1, \ldots, Q_{N_Y},P_{N_Y})$. 
Each collision is assumed to last for a small time $dt$ and is governed by a quadratic interaction Hamiltonian that we cast as $\mathcal{H} = \frac{1}{2} \hat{R}\trans H \hat{R}$ with $\hat{R} = (\hat{R}_X, \hat{R}_Y)$ and
\begin{equation}\label{H}
 \qquad H = \begin{pmatrix} H_X & C/\sqrt{dt} \\[0.2cm] C\trans/\sqrt{dt} & H_Y \end{pmatrix}.
\end{equation}
Here $H_X$ and $H_Y$ are the individual Hamiltonians of system and ancilla, and $C$ is the $N_X\times N_Y$ matrix accounting for the interaction between them. 
The scaling by $\sqrt{dt}$ is placed for convenience, as this yields simpler expressions in the limit of small $dt$~\cite{Rodrigues2019}. 

Gaussian states are completely characterized by the first moments $r = \langle \hat{R} \rangle$ and the covariance matrix 
$\sigma_{ij} = \frac{1}{2} \langle \{\hat{R}_i, \hat{R}_j \} \rangle - \langle \hat{R}_i \rangle\langle \hat{R}_j \rangle$. 
The ancillae are assumed to be prepared in Gaussian states with zero mean, $r_Y = 0$, and generic covariance matrix $\sigma_Y$. 
The system, on the other hand, is prepared with arbitrary $r_{X_0}$ and $\sigma_{X_0}$. 

By compounding different infinitesimal collisions, one can construct a continuous-time dynamics for the system~\cite{Wiseman1993,Wiseman1994,Doherty1999,Genoni2016}.
In Appendix~\ref{app:gaussian} we provide full details on this derivation, while here we only focus on the results. The unconditional dynamics is characterized by the matrices 
\begin{equation}\label{CXCY}
C_X = \Omega_X C, 
\qquad 
C_Y = \Omega_Y C\trans, 
\end{equation}
which, in general, are rectangular, with dimensions $N_X \times N_Y$ and $N_Y \times N_X$ respectively. 
Here $\Omega_X$ and $\Omega_Y$ are the symplectic forms with dimensions $N_X$ and $N_Y$. 
From $C_X$ and $C_Y$ we then define the drift and diffusion matrices
\begin{equation}
\label{gaussian_AAndD}
A = \Omega_X H_X + \frac{1}{2} C_X C_Y , \qquad
D = C_X \sigma_Y C_X\trans.
\end{equation}
Examples of typical system-ancilla interactions $C$,  as well as the  resulting shapes of $C_X$, $C_Y$, $A$ and $D$, are provided in Appendix~\ref{app:example_C}.
In the continuous time limit, one then finds that the first and second moments evolve according to the following linear equations in $r_X$ and $\sigma_X$
\begin{equation}\label{gaussian_uncond_1st_momentAndLyapunov}
\dot{r}_{X} = A r_{X},\qquad
\dot{\sigma}_{X} = A \sigma_{X} + \sigma_{X} A\trans + D, 
\end{equation}
which we refer to as a {\it Lyapunov problem}. 

The conditional dynamics, on the other hand, depends on two additional ingredients. 
The first is a functional of the covariance matrix
\begin{equation}\label{B_def}
B[\sigma_X] = \sigma_X C_Y\trans + C_X \sigma_Y.
\end{equation}
As discussed in Appendix~\ref{app:gaussian}, such functional encompasses the correlations developed between system and ancilla as a result of each collision. 
It is thus directly related to the information passed from the system to the ancillae. The second ingredient is the type of measurement performed on the state of the ancillae. 
We use here the framework of the so-called ``general-dyne'' measurements~\cite{Serafini2017}, whose outcomes are described by a random vector $z$ distributed according to a multivariate Gaussian with average given precisely by the final position of the ancillae $r_{Y'} =C_Y r_X \sqrt{dt}$ [cf. Appendix~\ref{app:gaussian}]. Such outcomes are thus directly proportional to the position of the system, but ``filtered'' by the matrix $C_Y$. 
Moreover, the covariance matrix of the outcomes $z$ is $\sigma_{Y} + \sigma_m$ with $\sigma_m$ being the covariance matrix of the noise induced by the specific choice of measurement. 
For a single-mode ancillary system with $\hat{R}_Y = (Q,P)$, a possible parametrization of such noise is~\cite{Genoni2016,Serafini2017,Belenchia2019}
\begin{equation}\label{gaussian_sigma_m}
\sigma_m = {\cal R}[\varphi]^T \begin{pmatrix} s/2 & 0 \\[0.2cm] 0 & 1/(2s) \end{pmatrix}{\cal R}[\varphi]+ \left(\frac{1-\eta}{\eta}+\Delta\right) I/2.
\end{equation}
The parameter $\eta\in [0,1]$ accounts for the detector efficiency, with $\eta = 1$ describing a perfectly efficient detector and $\eta = 0$ an inefficient one. Analogously, $\Delta\in[0,\infty)$ accounts for an additive Gaussian noise, 
while $s\in[0,\infty)$ defines the type of measurement being used:
$s = 0$ and $s = \infty$ correspond to homodyning $Q$ and $P$, respectively, while $s = 1$ is for a heterodyne measurement. Finally,  $R[\varphi]$ is a $2\times 2$ rotation matrix which allows us to describe general-dyne measurement on quadratures other than $Q$ and $P$.

With these  ingredients, we can now completely specify the conditional dynamics by defining the matrices 
\begin{equation}
\Lambda = C_Y\trans (\sigma_Y+\sigma_m)^{-1/2},\qquad
\Gamma = C_X \sigma_Y (\sigma_Y + \sigma_m)^{-1/2},
\end{equation}
as well as the functional

\begin{equation}\label{gaussian_chi}
\chi[\sigma] = B[\sigma] (\sigma_Y + \sigma_m)^{-1} B[\sigma]\trans
= (\sigma\Lambda + \Gamma)(\sigma\Lambda + \Gamma)\trans.
\end{equation}

The conditional first and second moments will then  evolve according to the {\it Riccati problem} (stochastic) 
\begin{equation}\label{gaussian_cond_1st_momentAndRiccati}
\begin{aligned}
dr_{X|\zeta} &= A r_{X|\zeta} dt + (\sigma_{X|\zeta}\Lambda + \Gamma) dw_t, \\
\dot{\sigma}_{X|\zeta} &= A \sigma_{X|\zeta} + \sigma_{X|\zeta} A\trans + D   - \chi[\sigma_{X|\zeta}],
\end{aligned}
\end{equation}
where $dw_t$ is a vector of independent Wiener increments satisfying $\langle dw \rangle = 0$ and $\langle dw\, dw\trans \rangle =\mathbb{I}_Y dt $. Eq.~(\ref{gaussian_cond_1st_momentAndRiccati}) shows that the first moments follow a dynamics induced by a stochastic Langevin equation, while the conditional covariance matrix evolves fully deterministically. This implies that
$\sigma_{X|\zeta}$ depends only on whether or not the measurement occurred and its nature, but not on the outcome $\zeta$. This peculiarity of Gaussian systems is responsible for a significant simplification in the formal description of the process, as it will soon be illustrated. 

The quantity $\chi[\sigma]$ in Eq.~\eqref{gaussian_chi} is often referred to as the \emph{innovation matrix} and represents the change in information from the measurement outcomes (recall that $B$ is associated with the system-ancilla correlations). 
For instance,  if $\eta \to 0$ in Eq.~\eqref{gaussian_sigma_m}, the matrix $\sigma_m$ diverges and hence $\chi \to 0$. 
The last two terms in Eq.~\eqref{gaussian_cond_1st_momentAndRiccati} thus represent a competition between the noise, accounted for by $D$, which tends to increase the modulus of the entries of $\sigma_{X|\zeta}$ (i.e., increasing they uncertainties), and the innovation $\chi$, which has the opposite effect.

\subsection{Information-theoretic and thermodynamic quantities}

The von Neumann entropy of an $N$-mode Gaussian system with covariance matrix $\sigma$ and positive symplectic eigenvalues $\{\nu_j\}$ is given by 
\begin{equation}\label{gaussian_vN}
S_\text{vN}(\sigma) = \sum\limits_{j=1}^N \left\{ \frac{\nu_j+1}{2} \ln \frac{\nu_j+1}{2} - \frac{\nu_j-1}{2} \ln \frac{\nu_j-1}{2}\right\}.
\end{equation}
Using the von Neumann entropy in the Gaussian case turns out to not be always very convenient, most remarkably because of the so-called {\it ultra-cold catastrophe}~\cite{Uzdin2019}, i.e. the divergence of thermodynamic quantities -- such as entropy production -- defined through $S_\text{vN}$ that is observed when the system of interest is affected by an environment prepared in a pure state. 
The reasons for such divergences can be traced back to the fact that the relative entropy $D(\rho'||\rho)$, which enters in the entropy production~\eqref{sigma}, diverges when the support of $\rho'$ is not contained in the support of $\rho$. 
Yet, such a situation is very common in quantum optical experiments where the ancillae entailed by our model would be embodied by the electromagnetic field of optical modes, which is {\it de facto} in its vacuum state~\footnote{In quantum optical experiments, the system of interest (cavity, mechanical mode, etc.) often interact with optical modes comprising a reservoir. The occupation number of optical frequency modes at room temperature is so small that justifies assuming that the mode is in the vacuum state, i.e. at $T=0$.}.

An alternative formulation for Gaussian systems is to use the Shannon entropy of the associated Wigner function~\cite{Santos2017b}. Such quantity, which for Gaussian states turns out to coincide with the  R\'enyi-2 entropy, takens on the particularly elegant form~\cite{Adesso2012} 
\begin{equation}\label{gaussian_renyi2}
S_2(\sigma) = \frac{1}{2} \ln |\sigma| + N \ln 2
\end{equation}
with $|\sigma|$ being the determinant of the covariance matrix.
The Wigner entropy is not affected by divergences, even when the environmental state is pure (e.g., a $T=0$ vacuum state modelling the interaction with an optical bath), and  converges to the von Neumann entropy in the classical limit of high temperatures.

Crucially, 
while in general the Wigner entropy does not enjoy a clear information-theoretic interpretation, its Gaussian version satisfies the strong-subadditivity inequality~\cite{Lieb1973,Adesso2012}, a key property for an entropy to acquire an information-theoretic sense, which legitimates our choice of entropic quantifier. 
Without affecting the generality of our conclusions, we henceforth take $S_2(\sigma)$ as our basic measure of entropy. 
Moreover, for simplicity we omit the constant offset $N\ln 2$ from Eq.~\eqref{gaussian_renyi2}.

Eq.~\eqref{gaussian_renyi2} provides a form for both $S(X_t)$ and $S(X_t|\zeta_t)$, the calculation of the latter being considerably simplified by the deterministic nature of the evolution of $\sigma_{X_t|\zeta_t}$ and its independence of $\zeta_t$. 
The Holevo information in Eq.~\eqref{holevo} takes the particularly simple form 
\begin{equation}\label{gaussian_I}
I(X_t\! : \! \zeta_t) = \frac{1}{2} \ln \frac{|\sigma_{X_t}|}{|\sigma_{X_t|\zeta_t}|},
\end{equation}
which -- through the identity $\frac{d}{dt} \ln |\sigma| = \tr (\sigma^{-1} \frac{d\sigma}{dt})$, and the Riccati problem in Eq.~\eqref{gaussian_cond_1st_momentAndRiccati} -- allows for the evaluation of the time-continuous information rate $\dot{I}$. We find 
\begin{equation}\label{gaussian_dIdt}
\dot{I} =  \frac{1}{2} \tr \left\{ \sigma_{X|\zeta}^{-1}~\chi[\sigma_{X|\zeta}]\right\}  - \frac{1}{2} \tr\left\{ \left( \sigma_{X|\zeta}^{-1} -\sigma_{X}^{-1} \right) D \right\},
\end{equation}
which should be split into a gain rate $\dot{G}$ and a loss rate $\dot{L}$, as in Eq.~\eqref{Delta_I_splitting}. A detailed derivation of such splitting is presented in Appendix~\ref{app:gaussian_stuff}, which shows that  
\begin{IEEEeqnarray}{rCl}
\label{gaussian_G2}
\dot{G} &=& \frac{1}{2} \tr \left\{ \sigma_{X|\zeta}^{-1}~\chi[\sigma_{X|\zeta}]\right\}, \\[0.2cm]
\label{gaussian_L2}
\dot{L} &=& \frac{1}{2} \tr\left\{ \left( \sigma_{X|\zeta}^{-1} -\sigma_{X}^{-1} \right) D \right\} .
\end{IEEEeqnarray}
These results are intuitive, as they demonstrate that $\dot{G}$ is associated with the innovation matrix $\chi$, while $\dot{L}$ depends on the noise encoded in $D$. Eqs.~\eqref{gaussian_dIdt}-\eqref{gaussian_L2} summarize the entire information dynamics in the Gaussian case.

We would like to conclude by remarking that the above calculations could also in principle be done using the von Neumann entropy in Eq.~\eqref{gaussian_vN}. We show in Fig.~\ref{fig:two_modes}(f) below that the results gathered through the Wigner and von Neumann entropy -- in the context of a specific example -- are nearly indistinguishable. 
However, the formal results obtained through the use of the von Neumann entropy are quite cumbersome, as they involve series expansions of the symplectic eigenvalues, which is a non-trivial task. Using the Wigner entropy therefore offers a significant simplification, making the interpretation of the results much clearer. 

\subsection{Conditions for the establishment of  ISS}

Eq.~\eqref{gaussian_G2} provides a clear condition for the existence of an ISS. 
Recalling the definition in Eq.~\eqref{gaussian_chi}, we can write $\dot{G}$ in the more symmetric form
\begin{equation}\label{gaussian_GBB}
\dot{G} = \frac{1}{2} \tr\Big\{ \sigma_{X|\zeta}^{-1/2} B[\sigma_{X|\zeta}] (\sigma_Y + \sigma_m)^{-1} B[\sigma_{X|\zeta}]\trans  \sigma_{X|\zeta}^{-1/2} \Big\}. 
\end{equation}
This is the trace of a positive semi-definite matrix
and both $\sigma_{X|\zeta}$ and $\sigma_Y + \sigma_m$ are quantum covariance matrices (which are therefore always strictly positive definite). Thus, 
\begin{equation}\label{gaussian_ISS_condition}
\dot{G} = 0, \qquad \text{iff}\qquad B[\sigma_{X|\zeta}] = 0. 
\end{equation} 
Of course, this assumes that the entries of the noise matrix $\sigma_m$ are finite; that is, that the measurement is not completely uninformative. 
For instance, if $\eta\to 0$ and/or $\Delta\to\infty$ in Eq.~\eqref{gaussian_sigma_m}, clearly we would have $\dot{G} = 0$ even if $B \neq 0$.

\subsection{Thermodynamic analysis}

Next we turn to the thermodynamics of the system. 
First, we evaluate the entropy flux, which in continuous-time takes the form of a rate
\begin{equation}
\label{gaussian_phi}
\dot{\Phi} =  \tr\left\{A\right\} +  \frac{1}{2} \tr\left\{ \sigma_{Y}^{-1} \tilde\sigma_{X} \right\}  
+ \frac{1}{2} \tilde r_{X}\trans  \sigma_{Y}^{-1}\tilde r_{X}, 
\end{equation} 
with $\tilde{r}_X=C_Y r_{X}$ and $\tilde\sigma_X=C_Y \sigma_{X} C_Y\trans$.
Similarly, the unconditional entropy production rate in Eq.~\eqref{2nd} becomes
\begin{equation}\label{gaussian_uncond_sigma}
\begin{aligned}
\dot{\Sigma}_u &= 2 \tr\{A\} +  \frac{1}{2} \tr\Big\{ \sigma_{X}^{-1} D + \sigma_{Y}^{-1} \tilde \sigma_X \Big\} 
+ \frac{1}{2} \tilde r_{X}\trans \sigma_{Y}^{-1} \tilde r_{X}\\
&=\dot{\Phi}+\tr\{A\}+\frac12\tr\left\{\sigma_X^{-1}D\right\}.
\end{aligned}
\end{equation}
Finally, 
we can compute the conditional entropy production $\dot{\Sigma}_c$ using Eq~\eqref{two_sigmas}
\begin{equation}
\label{gaussian_cond_sigma}
\begin{aligned}
\dot{\Sigma}_c &= 2\tr\{A\} + \frac{1}{2} \tr\Big\{ \sigma_Y^{-1} \tilde \sigma_X  + \sigma_{X|\zeta}^{-1}\big(D - \chi[\sigma_{X|\zeta}]\big) \Bigg\} +\frac{1}{2} \tilde r_{X}\trans \sigma_{Y}^{-1} \tilde r_{X}\\
&=\dot\Phi+\tr\{A\}+\frac12\tr\left\{\sigma_{X|\zeta}^{-1}\big(D - \chi[\sigma_{X|\zeta}]\big) \right\}.
\end{aligned}
\end{equation}
The last line in Eq.~\eqref{gaussian_cond_sigma} shows clearly that $\dot\Sigma_c$ coincides with the expression of $\dot\Sigma_{u}$ where $\sigma_X\to\sigma_{X|\zeta}$ and $D\to D-\chi[\sigma_{X|\zeta}]$, but without changing the associated entropy flux rate.
This, together with Eqs.~\eqref{gaussian_phi} and~\eqref{gaussian_uncond_sigma}, completely summarize the thermodynamics of continuous variable $\text{CM}^2$s.

%
%
\section{Examples and Applications}
\label{sec:example}
In this Section we illustrate the potential of the framework developed so far, by first tackling a paradigmatic example, and then moving to the modelling of an experiment in an optomechanical platform akin to the situation recently reported in Ref.~\cite{Rossi2019}.
%
%

%
%
\subsection{\label{ssec:two_mode}Example: two-mode ancilla}
%
%

\begin{figure*}
\includegraphics[width=\textwidth]{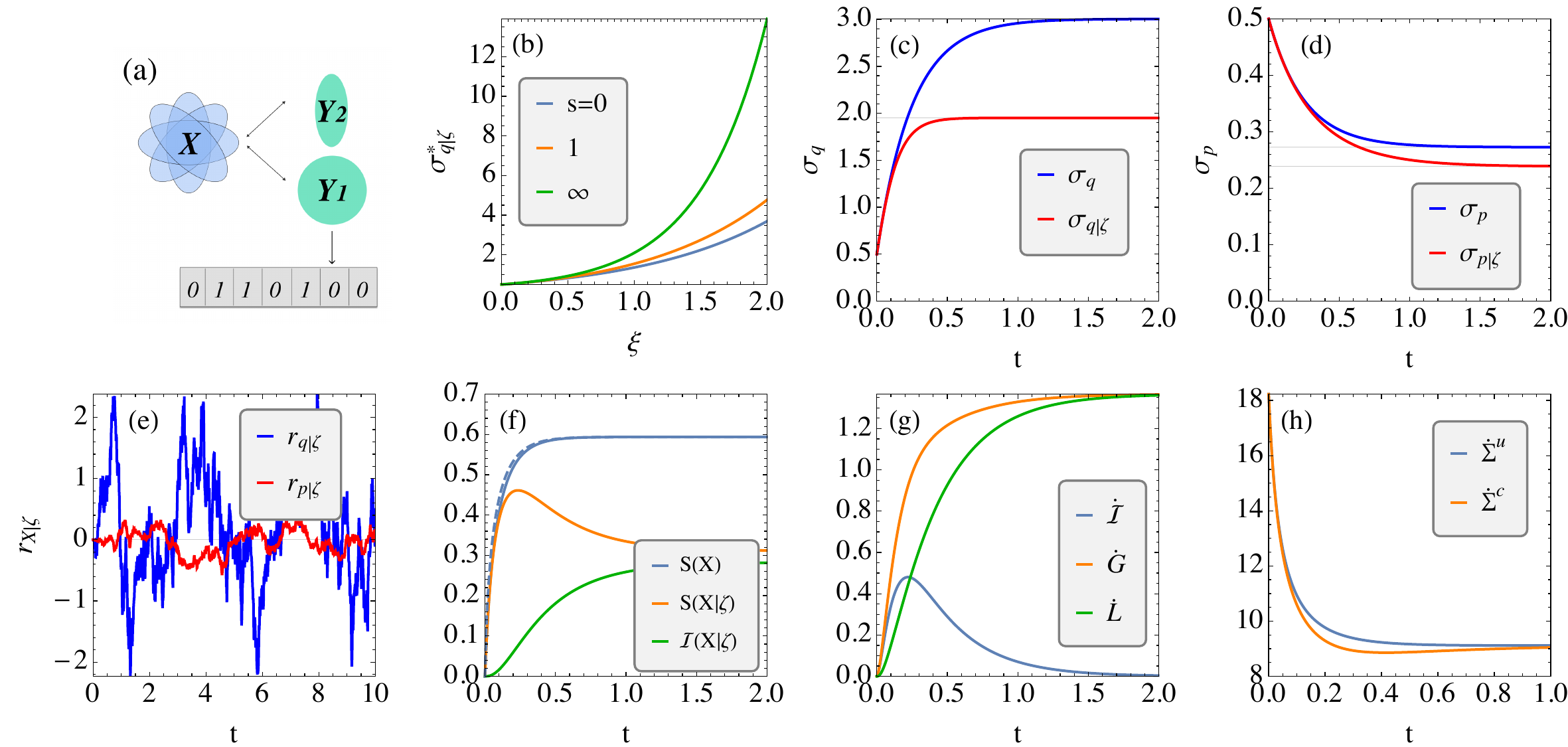}
\caption{
(a) $\text{CM}^2$ in the continuous-variable scenario, with a two-mode ancilla prepared in the state~\eqref{gaussian_example_sigma_y} and interacting with the beam-splitter matrix~\eqref{gaussian_beam_splitter_C}, with $\gamma = 1$. 
(b) Conditional steady-state variance of the $q$ quadrature for different measurement strategies, as a function of the squeezing  $\xi$ in the ancilla state: $s = 0$, homodyne in $Q$, $s = 1$, heterodyne and $s=\infty$ homodyne in $P$. 
For the latter, $\sigma_q^*$ coincides with the unconditional variance [Eq.~\eqref{gaussian_example1_uncond_sigma}]. 
(c), (d) Dynamics of the conditional and unconditional variances of $q$ and $p$, assuming the system starts in the vacuum, with $s = 1$ and $\xi = 1.2$.
(e) Sample trajectories of the conditional first moments. The choice of squeezing $\xi = 1.5$ in the ancillae cause $r_{q|\zeta}$ to fluctuate significantly more than $r_{p|\zeta}$. 
(f) Unconditional and conditional entropies, 
and Holevo information~\eqref{holevo}. The dashed line superimposed on the blue curve is the unconditional von Neumann entropy, scaled by a constant factor. For Gaussian states, it is nearly indistinguishable from the R\'enyi-2. 
(g) Information rate, $\dot{I}$, computed from Eq.~\eqref{gaussian_dIdt}, together with the information gain and loss rates $\dot{G}$ and $\dot{L}$ [Eqs.~\eqref{gaussian_G2} and~\eqref{gaussian_L2}].
The model presents an information steady-state, evidenced by a finite $\dot{G}$ even in the long-time limit. 
(h) Unconditional and conditional entropy production rates computed from Eqs.~\eqref{gaussian_uncond_sigma} and~\eqref{gaussian_cond_sigma}. 
}
\label{fig:two_modes}
\end{figure*}

We analyze a two-mode ancilla problem [cf. Fig.~\ref{fig:two_modes}(a)], where the first ancilla is prepared in its vacuum state ,while the second is in a squeezed  state of squeezing degree $\xi$. The covariance matrix of the environmental state is thus 
\begin{equation}\label{gaussian_example_sigma_y}
\sigma_Y = \begin{pmatrix}
{\mathbb I}/2 & {\mathbb O} \\
{\mathbb O} & {\mathbb S}/2 
\end{pmatrix}
\end{equation}
with ${\mathbb I}$ (${\mathbb O}$) the $2\times2$ identity (null) matrix and ${\mathbb S}={\rm diag}(e^{2\xi},e^{-2\xi})$.
The interest of this choice lies also in the fact that both ancillary sub-systems are here prepared in a pure state, which makes the standard formulation based on the von Neumann entropy inapplicable in light of the ultra-cold catastrophe. We assume the system interacts sequentially with each ancilla. 
Moreover, we take the interactions to be of an excitation-exchange type (cf. Appendix~\eqref{app:example:BS} for details), which results in a partial SWAP of the states of the colliding systems. 
The interaction matrix $C$ in Eq.~\eqref{H} takes the form
\begin{equation}\label{gaussian_beam_splitter_C}
C = \sqrt{2\gamma} \begin{pmatrix}
0 & -1 & 0 & -1 \\[0.2cm]
1 & 0  & 1 & 0 
\end{pmatrix},
\end{equation}
where $\gamma$ is the interaction strength. 
Finally, we assume only the first ancilla is measured. 
That is, we choose the measurement matrix $\sigma_m$ to be of the form [c.f. Eq.~\eqref{gaussian_sigma_m}]
\begin{equation}
\sigma_m = \frac{1}{2} \text{diag}\left( s, \frac{1}{s}, \frac{1-\eta}{\eta}, \frac{1-\eta}{\eta}\right).
\end{equation}
We then eliminate the information on the second ancilla by taking $\eta \to 0$. 

The unconditional steady-state is readily found by setting $\dot{\sigma}_X = 0$ in Eq.~\eqref{gaussian_uncond_1st_momentAndLyapunov}, which gives
\begin{equation}\label{gaussian_example1_uncond_sigma}
\sigma_X^* = 
\frac{{\mathbb S}+{\mathbb I}}{4}
\end{equation}
This is the average between the initial states of the two ancillae: The alternating collisions cause the system to homogenize to a state that is just the mean between the two states~\cite{Scarani2002}.

Similarly, we can also compute the conditional steady-state, by solving the equation $\dot{\sigma}_{X|\zeta} = 0$ in the Riccati problem of Eq.~\eqref{gaussian_cond_1st_momentAndRiccati}. 
The result is 
\begin{equation}\label{gaussian_example1_cond_sigma}
\sigma_{X|\zeta}^* = \frac{1}{2}\begin{pmatrix}
\sqrt{(1+s)(e^{2\xi}+ s)} -s & 0 \\[0.2cm]
0 & \frac{\sqrt{(1+s)(s e^{-2\xi}  + 1)}-1}{s}
\end{pmatrix}.
\end{equation}
The variance $\left(\sigma_{X|\zeta}^*\right)_{11}$ is shown in Fig.~\ref{fig:two_modes}(b) against the squeezing strength $\xi$ and for different measurement choices $s$ .
For $s\to \infty$, this tends to the unconditional value $\left(\sigma_q^* \right)_{11}= (e^{2\xi}+1)/4$, while for $s = 0$ it gives $e^{\xi}/2$.  For any value of $s$, we always have $\sigma_{q|\zeta}^* \leqslant \sigma_q^*$. 
Measuring therefore \emph{always} cools down both quadratures. However, the cooling performance depends on the type of measurement being considered.

Figs.~\ref{fig:two_modes}(c) and (d) shows the dynamics of the elements of the covariance matrix of $X$, which is assumed to be initially prepared in the vacuum state. 
Similarly, Fig.~\ref{fig:two_modes}(e) presents sample trajectories of the conditional first moments, $r_{q|\zeta}$ and $r_{p|\zeta}$. 
The results are for $s = 1$, which corresponds to performing a heterodyne measurement, so that the measurement is symmetric in both quadratures. 
However, the behavior of the $q$ and $p$ quadratures is fundamentally different. 
This is a consequence of the choice of initial ancilla state. We have chosen $\xi = 1.2$, meaning that $Y_2$ is squeezed in the $P$ direction (and hence expanded in the $Q$ direction). 
As a consequence, the steady-state covariance $\sigma_q$ is much larger than that of $p$, for both the conditional and unconditional dynamics. 
Interestingly, though, we also see that the cooling effect of measurement is  much more significant in the $q$ quadrature.

The results for the variances are reflected on both the information and thermodynamics of this example. 
In Fig~\ref{fig:two_modes}(f) we plot the unconditional and conditional entropies, as well as the Holevo information. The conditional entropy tends to a lower value than the unconditional one, in agreement with the findings for the variances. 
This happens because of the acquired information. 
Fig.~\ref{fig:two_modes}(g) shows the information rate $\dot{I}$, computed from Eq.~\eqref{gaussian_dIdt}. 
Initially a lot of information is obtained, but as time passes $\dot{I}$ tends to zero. 
However, the gain rate and loss rates, Eqs.~\eqref{gaussian_G2} and~\eqref{gaussian_L2}, tend to a finite value in the steady-state, thus characterizing an ISS. 
Finally, a comparison between the unconditional and conditional evolution of the entropy production is shown in Fig.~\ref{fig:two_modes}(h), where it can be seen that both decay from an initially high value towards a non-zero steady-state value. 
The reason why the entropy production rate is initially high is because the initial state of the system is very far from equilibrium. Also, the difference $\sigma^c - \sigma^u$ is larger for intermediate times, which is when $\dot{I}$ is the largest. At $t= 0$ and at $t = \infty$, both quantities coincide, as they should. 

The results of Fig.~\ref{fig:two_modes} clearly show that the system tends to an ISS. 
According to Eq.~\eqref{gaussian_ISS_condition}, the condition for this to be the case is to have $B[\sigma_{X|\zeta}^*] > 0$. 
In our case, using Eqs.~\eqref{gaussian_example_sigma_y}, \eqref{gaussian_beam_splitter_C} and~\eqref{gaussian_example1_cond_sigma}, we find 
\[
B[\sigma_{X|\zeta}^*] = \sqrt{\frac{\gamma}{2}} \begin{pmatrix}
1-2\sigma_{Q|\zeta}^* & 0 & e^{2\xi} - 2 \sigma_{Q|\zeta}^* & 0 \\[0.2cm]
0 & 1- 2 \sigma_{P|\zeta}^* & 0 & e^{-2\xi} - 2 \sigma_{P|\zeta}^*
\end{pmatrix}.
\]
where $\sigma_{Q|\zeta}^*$ and $\sigma_{P|\zeta}^*$ are the entries of Eq.~\eqref{gaussian_example1_cond_sigma}. 
We therefore see the conditions for the existence of an ISS are quite light. 
Essentially, as long as the steady-state of the system is neither that of $Y_1$ nor that of $Y_2$, information will continue to be acquired in every collision. 
This result also provides guidelines on how different measurement strategies affect the ISS. 
For instance, suppose we were to measure ancilla $Y_2$ instead of $Y_1$. 
From Eq.~\eqref{gaussian_GBB} we have that $\dot{G} \propto B (\sigma_Y + \sigma_m)^{-1} B\trans$ and 
measuring $Y_2$ means introducing an infinite amount of noise in the $Y_1$ block of $\sigma_m$. 
This would then eliminate the left block of $B$.

\subsection{Global~vs.~reduced dynamics}

We now use the continuous variable results to make a small digression about an important point in quantum and stochastic thermodynamics. 
Consider a general scenario of a system interacting with a bath. 
Very often, this process is described by an effective reduced description, such as a quantum master equation. 
The point we wish to address is that, while this description may be adequate for describing the dynamics, it does not necessarily suffice to describe the \emph{thermo}dynamics~\cite{Landi2020a}. 
This can be illustrated by the following minimal example. 
Consider a single mode subject to a standard thermal bath, as described by the Gorini-Kossakowski-Lindblad-Sudarshan (GKLS) master equation
\begin{IEEEeqnarray}{rCl}\label{qubit_M_equation}
\frac{d\rho}{dt} 
&=& - i [H, \rho] + \mathcal{D}(\rho) \\[0.2cm]  
&=& - i [H, \rho] + \gamma(\bar{n}+1) {\cal D}_- + \gamma \bar{n} {\cal D}_+,   \nonumber
\end{IEEEeqnarray}
where 
$H = \omega a^\dagger a$,
${\cal D}_- = a\rho a^\dagger - \frac{1}{2}\{a^\dagger a, \rho\}$ and 
${\cal D}_- = a^\dagger\rho a - \frac{1}{2}\{a a^\dagger, \rho\}$.
Moreover, 
$\gamma>0$ is the dissipation rate, and $\bar{n} = (e^{\omega/T}-1)^{-1}$ is the mean number of excitations in the bath. 
In this scenario, one would naturally associate the heat current to the bath with 
\begin{equation}
\frac{d\langle H \rangle}{dt} = \tr\Big\{ H \mathcal{D}(\rho) \Big\}. 
\end{equation}
In the long-time limit the system will tend to an equilibrium state, characterized by a zero current. 
However,  suppose instead that the same system is coupled to two baths, at different temperatures $T_1$ and $T_2$. 
In a weak-coupling approximation, the resulting master equation for the two baths will be additive, so that we would simply have 
\begin{equation}\label{qubit_M_equation_2}
\frac{d\rho}{dt} = - i [H, \rho] + \mathcal{D}_1(\rho) + \mathcal{D}_2(\rho), 
\end{equation}
where each dissipator $\mathcal{D}_i$ is defined as in Eq.~\eqref{qubit_M_equation}, with parameters $\gamma_i$ and $\bar{n}_i$. 
Eq.~\eqref{qubit_M_equation_2} can be recast in a form involving a single dissipator by defining $\mathcal{D}' = \mathcal{D}_1 + \mathcal{D}_2$, which would be of the form of Eq.~\eqref{qubit_M_equation} with parameters $\gamma' = \gamma_1+\gamma_2$ and $\bar{n}' = (\gamma_1 \bar{n}_1 + \gamma_2 \bar{n}_2)/(\gamma_1 + \gamma_2)$.  While such reformulation would suggest thermalization with a single bath at the effective temperature $T'=\omega/\ln(1+1/\bar{n}')$, in reality, the process itself is clearly different and would drive the system to a non-equilibrium steady state. This is seen from the fact that, as  long as $T_1 \neq T_2$, we will have $\tr\big\{ H \mathcal{D}_i \big\} \neq 0$, meaning there will be a current of heat from one bath to the other. 
However, this can only be observed if one has access to the additional information that $\mathcal{D}' = \mathcal{D}_1 + \mathcal{D}_2$ and the local currents. A rigorous thermodynamic description therefore requires that one properly identifies all possible heat sources. 
Despite its simplicity, this example illustrates well how the thermodynamic interpretation can be \emph{fundamentally} altered depending on the amount of global information one has access to. 
Note also that this is not a quantum feature, as the same problem also appears in stochastic thermodynamics, as discussed in detail, e.g. in Ref.~\cite{Esposito2010e}. 

We will now analyze this issue from the viewpoint of continuous variable models. 
The reduced dynamics of the system is specified by the four matrices  $A, D, \Gamma$ and $\Lambda$, while the full global dynamics is also specified by $\sigma_Y$, $C_X$ and $C_Y$. 
Any property that can be expressed solely as a function of  the former set of matrices can thus be found from the reduced dynamics alone. 
Let us then analyze the entropy flux rate Eq.~\eqref{gaussian_phi} from this perspective. 
We can rewrite the flux rate in terms of $\Upsilon = C_Y\trans \sigma_Y^{-1} C_Y$ as
\[
\dot{\Phi} = \tr\{A\} + \frac{1}{2} \tr\{\Upsilon \sigma_X\} + \frac{1}{2} r_X\trans \Upsilon r_X. 
\] 
In general $\Upsilon$ cannot be constructed solely from $A, D, \Gamma, \Lambda$, which shows the global character of the flux.
This becomes quite important in light of the additivity property of the flux rate stated in Eq.~\eqref{phi_multiple}. 
To see this, suppose $Y$ has multiple, initially independent, internal units, so that 
$\sigma_Y = \bigotimes_j\sigma_{Y_j}$. 
This entails
\begin{equation}
C_X = 
\begin{pmatrix} C_{X_1} & C_{X_2} & \ldots  
\end{pmatrix}, 
\qquad 
C_Y = 
\begin{pmatrix}  C_{Y_1}\\[0.2cm]  C_{Y_2} \\[0.2cm]  \vdots 
\end{pmatrix}, 
\end{equation}
and as a consequence 
$\Upsilon = \sum_j C_{Y_j}\trans \sigma_{Y_j}^{-1} C_{Y_j} = \sum_j \Upsilon_j$, 
We also define the dissipative part  of matrix $A$ as $A_d= \frac{1}{2} C_X C_Y\sum_j A_{dj}$ (with $A_{dj} = \frac{1}{2} C_{X_j} C_{Y_j}$), which is the only part  contributing to $\tr\{A\}$. 
The flux rate is thus additive, and reads
\begin{equation}\label{gaussian_phi_sum_0}
\dot{\Phi}= \sum_j\Phi_j=\frac12 \sum\limits_j \left[\tr\{C_{X_j} C_{Y_j}\} + \frac{1}{2} \tr\{\Upsilon_j \sigma_X\} + \frac{1}{2} r_X\trans \Upsilon_j r_X \right].
\end{equation}
Each term in the sum identifies the entropy flux rate to the individual ancillae and, hence, to each independent source of dissipation within the system. In the particular case where the matrices $C_{X_j}$ are invertible, we can also relate $\Upsilon$ with the diffusion matrix $D$ [cf. Eq.~\eqref{gaussian_AAndD}], thus giving the alternative decomposition of the flux rate  as  
\begin{equation}\label{gaussian_phi_sum}
\dot\Phi = \sum\limits_j  \left[ \tr\{A_{dj}\} + 2 \tr\left\{A_{dj}\trans D_j^{-1} A_{dj} \sigma_X\right\} + 2~r_X\trans A_{dj}\trans D_j^{-1} A_{dj} r_X\right]. 
\end{equation}
This expression preserves the correct identification of the dissipation channels. 
Note, however, that it requires not only $A_d$ and $D$, but also their specific decompositions in terms of $A_{dj}$ and $D_j$. 
Finally, we also mention that the flux rate can only be expressed in the form in Eq.~\eqref{gaussian_phi_sum} if the matrices $C_{X_j}$ are invertible. There are many cases when this does not hold true, as illustrated for instance in Appendix~\ref{app:example:QQ}. 
In those cases, one must rely on the original expression Eq.~\eqref{gaussian_phi_sum_0}, which holds for any interaction matrix.

\subsection{Modeling an optomechanical experiment}

Finally, we  employ our framework to describe the experiment performed in Ref.~\cite{Rossi2019}. 
The setup consists of an intra-cavity mechanical mode embodied by  a vibrating membrane, subjected to two external baths. The first is a standard thermal bath, associated with a phononic background for the mechanical mode. 
The second bath is optical, and provided by  the field of the cavity, which is eliminated adiabatically from the dynamics of the system and, by being continuously monitored, gives information on the mechanical system. 
The scenario is therefore similar in spirit to the two-mode example of Fig.~\ref{fig:two_modes}.

The evolution of the system is described by the  stochastic master equation~\cite{Szorkovszky2011,Doherty2012}
\begin{equation}\label{Coppe_drho}
d\rho = \mathcal{L}_\text{th-qba} dt + \sqrt{\eta \Gamma_\text{qba}} \bigg( \mathcal{H}[q] dw_1 + \mathcal{H}[p] dw_2\bigg), 
\end{equation}
where 
\begin{IEEEeqnarray}{rCl}
\mathcal{L}_\text{th-qba} &=& (\Gamma_m (\bar{n}+1)+\Gamma_\text{qba}) {\cal D}_- + (\Gamma_m \bar{n}+\Gamma_\text{qba}){\cal D}_+, 
\end{IEEEeqnarray}
represents the Lindblad dissipator, including the quantum back-action mechanism induced by the monitored optical baths, with ${\cal D}_\pm$
defined below Eq.~\eqref{qubit_M_equation}.
In Eq.~(\ref{Coppe_drho}) we also defined $\mathcal{H}[\mathcal{O}] =  \mathcal{O} \rho+ \rho \mathcal{O}^\dagger - \rho \tr((\mathcal{O}+\mathcal{O}^\dagger)\rho)$, which is associated with the continuous measurements, with $\eta \in [0,1]$ denoting the measurement efficiency.
Finally, $dw_1$ and $dw_2$ are two independent Wiener increments. 

Eq.~(\ref{Coppe_drho}) leads to unconditional and conditional dynamics described  by Eqs.~(\ref{gaussian_uncond_1st_momentAndLyapunov}) and (\ref{gaussian_cond_1st_momentAndRiccati}), with 
\begin{equation}
\label{Coppe_matrices}
A = -\frac{\Gamma_m}{2} {\mathbb I},\, 
D = \left[ \Gamma_m \left(\bar{n}+\frac12\right) + \Gamma_\text{qba}\right] {\mathbb I},\, 
\chi[\sigma] = 4 \eta \Gamma_\text{qba} \sigma_X^2. 
\end{equation} 
These results are intuitive: the diffusion matrix $D$, which is responsible for the noise, is associated to both the thermal and optical baths. The innovation matrix $\chi$, on the other hand, is associated only to the optical bath, which is the only one being measured. Moreover, $\chi \propto \eta$, so that if $\eta = 0$ (fully inefficient measurement), the innovation vanishes. 
Slightly less intuitive is the fact that the optical bath does not affect the damping matrix $A$. This is associated to the specific way in which the optical mode couples to the mechanical system. 

In order to be able to properly describe the thermodynamics of this system, we must now construct a $\text{CM}^2$ which reproduces the matrices in Eqs.~\eqref{Coppe_matrices} at the level of the reduced dynamics, looking for 
its minimal construction.
First, it is worth stressing that the thermal and optical baths are independent. The former does not have to be modelled by a collisional model as it is not monitored. Its effects could thus be described by a master equation. In order to better match with the notation of the remainder of the paper, however, we shall assume that the thermal part is described by a collisional model as well. 
In this case, it can be generated by using a single-mode thermal ancilla interacting with the system via an excitation-exchange interaction, as studied in Sec.~\ref{ssec:two_mode}. 
The description of the optical mode, on the other hand, is less trivial. 
A special feature of the stochastic master equation Eq.~\eqref{Coppe_drho} is that it allows one to independently monitor the two mechanical quadratures $q$ and $p$. 
The optical bath must, therefore, be itself composed of at least two modes. 
Hence, the ancilla in this model must have a total of 3 modes, with initial state 
\begin{equation}
\sigma_Y = 
\begin{pmatrix}
(\nn){\mathbb I}&{\mathbb O}&{\mathbb O}\\
{\mathbb O}&{\mathbb I}/2&{\mathbb O}\\
{\mathbb O}&{\mathbb O}&{\mathbb I}/2
\end{pmatrix}.
\end{equation}
The first ancillary mode is in a thermal state with occupation number $\bar{n}$, while the optical ancillae are initially in the vacuum state. 

Next we turn to the interaction matrices for the optical baths. 
The peculiar feature of this interaction is that it generates no contribution to the damping matrix $A$ in Eq.~\eqref{Coppe_matrices}. 
As discussed in Appendix~\ref{app:example:QQ}, this feature is generated by position-position or momentum-momentum couplings: the second ancilla (i.e. the first optical one) is used to monitor $q$ via the interaction Hamiltonian  
$-\sqrt{2\Gamma_\text{qba}} q Q_2$, while the third ancilla monitors the mechanical momentum $p$ via the term $\sqrt{2\Gamma_\text{qba}} p P_3$.
The interaction matrix $C$ will thus have the form 
\begin{equation}
C = \begin{pmatrix}
0 & -\sqrt{\Gamma_m} & -\sqrt{2\Gamma_\text{qba}} & 0 & 0 & 0 \\[0.2cm]
\sqrt{\Gamma_m} & 0 & 0 & 0 & 0 & \sqrt{2 \Gamma_\text{qba}} \end{pmatrix}.
\end{equation}
This  leads to the matrices [cf. Eq.~(\ref{CXCY})]
\begin{IEEEeqnarray}{rCl}
C_X &=& \begin{pmatrix}
\sqrt{\Gamma_m}{\mathbb I} & \sqrt{2\Gamma_\text{qba}}\sigma_- & \sqrt{2\Gamma_\text{qba}}\sigma_-\end{pmatrix},\label{CxMatrix}\\[0.2cm]
C_Y &=& \begin{pmatrix}
- \sqrt{\Gamma_m}{\mathbb I} \\[0.2cm]
\sqrt{2\Gamma_\text{qba}}\sigma_- \\[0.2cm]
 \sqrt{2\Gamma_\text{qba}}\sigma_+
 \end{pmatrix}.
\end{IEEEeqnarray}
The drift and diffusion matrices Eqs.~\eqref{gaussian_AAndD} will then be given by Eq.~\eqref{Coppe_matrices}. 

Finally, to reproduce the innovation matrix in Eq.~\eqref{Coppe_matrices}, the measurement matrix $\sigma_m$ must have the form $\sigma_m = \bigoplus^3_{j=1}\sigma_{m_j}$, where each $\sigma_{m_j}$ is given by Eq.~\eqref{gaussian_sigma_m}, with the following parameters: 
First, for the thermal ancilla $\eta_1 = 0$ with $s_1$ arbitrary. 
Then, since the first optical ancilla is coupled through a position-position mechanism to the mechanical system, we actually have to detect its momentum $P_2$. That is, we set $\eta_2 = \eta$ and $s_2 = \infty$. Finally, for the second optical ancilla, we set $\eta_3 = \eta$ and $s_3 = 0$, so that we homodyne $Q_3$. 
With these choices for the matrix $\sigma_m$, one then reproduces exactly the innovation matrix in Eq.~\eqref{Coppe_matrices}. The minimal CM$^2$ that we have deduced here can consistently describe the dynamics and thermodynamics of the experiment performed in Refs.~\cite{Rossi2019,Rossi2020}. 

In particular, in Ref.~\cite{Rossi2020}, the analysis of the thermodynamics of the system addressed here was presented by working at the level of the system master equation. However, in order to obtain the correct expression of the entropy fluxes, a refined analysis was needed, starting from the full mechanical-cavity system master equation before the adiabatic elimination of the latter could be performed~\cite{Szorkovszky2011,Doherty2012} leading to Eq.~\eqref{Coppe_drho}. Here we can clearly see why that was the case. Indeed, we can see from Eq.~\eqref{CxMatrix} that the matrices $C_{X_i}$ for $i=2,3$ --- which reproduce the dynamics as described by Eq.~\eqref{Coppe_drho} obtained after adiabatic elimination and other approximations --- are singular. Following our previous discussion, the entropy flux in this case cannot be written solely in terms of the reduced dynamics. This  strengthen further the point we made in the previous section on the importance of global information for the consistent description of the thermodynamics of the system.

%
%
\section{\label{sec:conc} Conclusions}
%
%
We have investigated the thermodynamics of continuously monitored quantum Gaussian systems, reformulating the problem in a collisional model framework~\cite{firstpaper}. By focusing on continuous variables, we were able to connect the entropy production and flux rates, characterizing the thermodynamics of the quantum process, to purely informational quantities and highlight the role of information in characterizing the irreversibility of the dynamics and sustaining non-trivial asymptotic states, i.e., the ISSs. 

Our work also advances the results first presented in~\cite{Belenchia2019}, where a semiclassical, phase-space description of continuously measured Gaussian systems thermodynamics was presented. In particular, we have addressed here the issue of how much global knowledge is required to properly describe the thermodynamics of Gaussian systems. Indeed, the dynamics of an open quantum system is determined by the knowledge of its master equation. This, in turn, requires certain information on the environment to which the system is exposed. However, this information is in general not sufficient to properly characterise the thermodynamics of the system, as exemplified by the experiment in~\cite{Rossi2020}. We have shown that the collisional model framework allows to identify the cases in which global information beyond the master equation is needed in order to have a consistent thermodynamic description of the quantum process of interest.   

The formalism we have developed in this work is applicable to the wide class of Gaussian systems and processes which are of pivotal importance in quantum information science and quantum technologies. 
In fact, we also provided an account of a recent experiment in optomechanics, showing that our formalism should be useful in describing a broad variety of quantum-coherent experiments.

\acknowledgements
We acknowledge support from the Deutsche Forschungsgemeinschaft (DFG, German Research Foundation) project number BR 5221/4-1, the  MSCA project pERFEcTO (Grant No. 795782), the H2020-FETOPEN-2018-2020 TEQ (grant nr. 766900), the DfE-SFI Investigator Programme (grant 15/IA/2864), COST Action CA15220, the Royal Society Wolfson Research Fellowship (RSWF\textbackslash R3\textbackslash183013), the Leverhulme Trust Research Project Grant (grant nr.~RGP-2018-266), the UK EPSRC (grant nr.~EP/T028106/1).

%
\appendix

%
%
\section{\label{app:gaussian} Construction of the Gaussian $\text{CM}^2$}
%
%

In this appendix we detail the derivation of the main results of Sec.~\ref{ssec:GaussianCM2}. 
We begin by focusing on a single collision described by the Hamiltonian~\eqref{H}.
The Heisenberg evolution of the quadratures  after a time $dt$ is given by $\hat{R}(dt) = e^{i \Omega_{XY} H dt} \hat{R}(0)$, where $\Omega_{XY}$ is the symplectic form of dimensions $2(N_X + N_Y)$. 
Expanding for small $dt$, we then find 
\begin{IEEEeqnarray}{rCl}
\label{gaussian_RX}
\hat{R}_{X'} &=&  \hat{R}_X + A \hat{R}_X dt + C_X \hat{R}_Y \sqrt{dt} , \\[0.2cm]
\label{gaussian_RY}
\hat{R}_{Y'} &=&  \hat{R}_Y + A_Y \hat{R}_Y dt + C_Y \hat{R}_X \sqrt{dt} , 
\end{IEEEeqnarray}
where the matrix $A$ is defined in Eq.~\eqref{gaussian_AAndD}. 
In addition, we also defined 
\begin{equation}
\label{gaussian_AY}
A_Y =\Omega_Y H_Y + \frac{1}{2} C_Y C_X. 
\end{equation}
From Eqs.~\eqref{gaussian_RX} and~\eqref{gaussian_RY} we then find that the first moments, after the collision, are given by 
\begin{IEEEeqnarray}{rCl}
\label{gaussian_rX}
r_{X'} &=&  r_X + A r_X dt , \\[0.2cm]
\label{gaussian_rY}
r_{Y'} &=& C_Y r_X \sqrt{dt} , 
\end{IEEEeqnarray}
The ancilla is displaced by an amount proportional to $r_{X}$, but this is ``filtered'' by $C_{Y}$, which can cause the ancilla to become blind to some of the system's quadratures. 
This becomes particularly clear from the examples discussed in Appendix~\ref{app:example_C}.

Similarly, we can look at the evolution of the second moments.
We parametrize the covariance matrix after the collision as 
\begin{equation}
\sigma_{X'Y'} = \begin{pmatrix} \sigma_{X'} & \xi_{X'Y'} \\[0.2cm]
\xi_{X'Y'}\trans & \sigma_{Y'}
\end{pmatrix}.
\end{equation}
Eqs.~\eqref{gaussian_RX} and~\eqref{gaussian_RY} then yield
\begin{IEEEeqnarray}{rCl}
\label{gaussian_sigmaX}
\sigma_{X'} 
&=& \sigma_X + \Big(A \sigma_X + \sigma_X A\trans + D \Big) dt , 
\\[0.2cm]
\label{gaussian_sigmaY}
\sigma_{Y'} &=& \sigma_Y + \big(A_Y \sigma_Y + \sigma_Y A_Y\trans + C_Y \sigma_X C_Y\trans \Big) dt, 
\\[0.2cm]
\nonumber
\xi_{X'Y'} &=& (\sigma_X C_Y\trans + C_X \sigma_Y) \sqrt{dt} \\[0.2cm]
&:=& B[\sigma_X] \sqrt{dt}.
\label{gaussian_xiXY}
\end{IEEEeqnarray}
where $B$ was defined in Eq.~\eqref{B_def} and $D$ is the diffusion matrix, defined in Eq.~\eqref{gaussian_AAndD}.

On the other hand, the conditional state of the system, given a certain measurement outcome,  is still Gaussian, with first and second moments given by \cite{Genoni2016}
\begin{IEEEeqnarray}{rCl}
\label{gaussian_rXZ}
r_{X'|z} &=& r_{X'} + \xi_{X'Y'} (\sigma_{Y'} + \sigma_m)^{-1}(z - r_{Y'}), \\[0.2cm]
\label{gaussian_sigmaXZ}
\sigma_{X'|z} &=& \sigma_{X'} - \xi_{X'Y'} (\sigma_{Y'} + \sigma_m)^{-1} \xi_{X'Y'}\trans. 
\end{IEEEeqnarray}
Conditioning  updates the average by a term proportional to the correlations $\xi_{X'Y'}$, as well as the outcomes $z$. 
Since $z$ is random,  $r_{X'|z}$ will be stochastic.
The covariance matrix $\sigma_{X'|z}$, on the other hand, is reduced by the presence of the 2nd term in~\eqref{gaussian_sigmaXZ}, called the Schur complement.
Note that this term is, by construction, positive semi-definite, so that indeed conditioning always reduces the  uncertainty about the system, as expected. 

Eqs.~\eqref{gaussian_rXZ} and~\eqref{gaussian_sigmaXZ} are general, in that they do not require the collision time to be infinitesimal. 
On the other hand, expanding in powers of $dt$ and using Eqs.~\eqref{gaussian_rX}-\eqref{gaussian_xiXY}, we find  
\begin{IEEEeqnarray}{rCl}
\nonumber
r_{X'|z} &=& r_{X'} + B[\sigma_X] (\sigma_Y + \sigma_m)^{-1/2} dw \\[0.2cm]
\label{gaussian_rXZ_2}
&=& r_{X} + A r_X dt  + B[\sigma_X] (\sigma_Y + \sigma_m)^{-1/2} dw, \\[0.2cm]
\nonumber
\sigma_{X'|z} &=& \sigma_{X'} - \chi[\sigma_X] dt \\[0.2cm]
\label{gaussian_sigmaXZ_2}
&=& \sigma_{X}  + \Big(A \sigma_X + \sigma_X A\trans + D   - \chi[\sigma_X]\Big) dt , 
\end{IEEEeqnarray}
where $dw = (\sigma_Y + \sigma_m)^{-1/2} (z - r_{Y'})$ can be shown to behave as a Wiener white noise term (that is, $\langle dw \rangle = 0$ and $\langle dw dw\trans \rangle = dt\,\mathbb{I}_Y$).  

The first line in~\eqref{gaussian_sigmaXZ_2} can be viewed as a manifestation of the so-called \emph{law of total variance}~\cite{Bickel1977}. 
In classical probability theory, the variance of a random variable $X'$ can be written as 
\begin{equation}\label{Eve}
\Var(X') = \E_z\bigg( \Var(X'|z) \bigg) + \Var_z\bigg( \E(X'|z) \bigg).
\end{equation}
The first term, called the \emph{within-group} variation, measures the fluctuations $\Var(X'|z)$ within a given outcome $z$ (i.e., within a given ``group''),  and then averaged it over all outcomes. 
Conversely, the second term, called \emph{between-groups},  quantifies how much the conditional average $\E(X'|z)$ fluctuates between different outcomes $z$. 

The law naturally extends for covariance matrices. 
Then, since we only condition on classical random variables $z$, the logic remains true, even though the system is quantum. 
The within group term is thus $\E_z\big(\sigma_{X'|z}\big)$.
However, since $\sigma_{X'|z}$ doesn't  depend on $z$, this simplifies to $\E_z\big(\sigma_{X'|z}\big) = \sigma_{X'|z}$.
Similarly, the between groups is $\text{Cov}_z(r_{X'|z})$, where $\text{Cov}$ stands for the covariance matrix of the random vector $r_{X'|z}$. 
Hence, by comparison, moving  $\chi[\sigma_{X}]dt$ to the left of Eq.~\eqref{gaussian_sigmaXZ_2}, we  see 
that the between group contribution is precisely 
\begin{equation}
\Var_z (r_{X'|z}) = \chi[\sigma_X] dt. 
\end{equation}
This provides another neat interpretation to the innovation matrix: 
It describes how $r_{X'|z}$ fluctuates between different outcomes $z$.

As a technical note, we mention that one could also, in principle, write down equations for the conditional state of the ancilla, given the measurement outcomes.
That is, $r_{Y'|z}$ and $\sigma_{Y'|z}$.
This, however, is not so easy, for it requires knowledge of the exact generalized measurement operators $M_z$. 
The noise covariance matrix $\sigma_m$, we are using here, only specifies  the resulting POVM.
And there is an infinite number of non-trivial choices of generalized measurements which yield the same POVM. 
Luckily, all quantities, both informational and thermodynamic, can be expressed without knowledge of $r_{Y'|Z}$ and $\sigma_{Y'|Z}$, as will be shown below.

Having established the evolution rules for a single collision, it is now straightforward to compound them and construct the continuous time dynamics.
The unconditional dynamics, for instance, is given by the update rules~\eqref{gaussian_rX} and~\eqref{gaussian_sigmaX} which, when adapted to multiple collisions, become:
\begin{IEEEeqnarray}{rCl}
r_{X_t} &=& r_{X_{t-1}} + A r_{X} dt. 
\\[0.2cm]
\sigma_{X_t} &=& \sigma_{X_{t-1}} + \Big(A \sigma_{X_{t-1}} + \sigma_{X_{t-1}} A\trans + D \Big) dt , 
\label{gaussian_uncond_sigma_discrete}
\end{IEEEeqnarray}
Dividing by $dt$ on both sides and taking the limit $dt\to 0$ then yields precisely Eqs.~\eqref{gaussian_uncond_1st_momentAndLyapunov}.

For the conditional dynamics, some care must be taken. 
Eqs.~\eqref{gaussian_rXZ_2} and~\eqref{gaussian_sigmaXZ_2} refer to a single collision. 
Hence, the quantities $r_X$ and $\sigma_X$ that appear on the right-hand side, are actually the state of the system before that collision. 
In the case of a conditional dynamics, this would then be $r_{X_{t-1}|\zeta_{t-1}}$ and $\sigma_{X_{t-1}|\zeta_{t-1}}$. 
The left-hand side will then be associated with  $X_{t} | \zeta_{t}$. 
Thus, the conditional evolution will be described by 
\begin{widetext}
\begin{IEEEeqnarray}{rCl}
\label{gaussian_rXZ_t}
\IEEEeqnarraynumspace
r_{X_t|\zeta_t} &=& r_{X_{t-1}|\zeta_{t-1}} + A r_{X_{t-1}|\zeta_{t-1}} dt  + B[\sigma_{X_{t-1}|\zeta_{t-1}}] (\sigma_Y + \sigma_m)^{-1/2} dw_t, \\[0.2cm]
\label{gaussian_sigmaXZ_t}
\sigma_{X_t|\zeta_t} &=& \sigma_{X_{t-1}|\zeta_{t-1}}  + \Big(A \sigma_{X_{t-1}|\zeta_{t-1}} + \sigma_{X_{t-1}|\zeta_{t-1}} A\trans + D   - \chi[\sigma_{X_{t-1}|\zeta_{t-1}}]\Big) dt . 
\IEEEeqnarraynumspace
\end{IEEEeqnarray}
\end{widetext}
Eq.~\eqref{gaussian_rXZ_t} leads to the Langevin equation in~\eqref{gaussian_cond_1st_momentAndRiccati}, while Eq.~\eqref{gaussian_sigmaXZ_t}, when taking the limit $dt \to 0$, leads to the Riccati equation in~\eqref{gaussian_cond_1st_momentAndRiccati}.

%
%
\section{\label{app:example_C} Examples of system-ancilla interactions in the continuous-variable case}
%
%

In this appendix, we provide examples of some typical system-ancilla interactions in the continuous-variable scenario. 
We also discuss the basic structure of the resulting matrices $C_X$, $C_Y$ in Eq.~\eqref{CXCY}, as well as the matrices $A$ and $D$ in Eqs.~\eqref{gaussian_AAndD}, which  enter in many of the equations in Sec.~\ref{ssec:GaussianCM2}.

\subsection{\label{app:example:BS}Quantum-optical master equation}

Suppose the system and ancilla are each comprised of a single mode of radiation, described by annihilation operators $a_X$ and $a_Y$ and interacting with a beam-splitter Hamiltonian 
\begin{equation}\label{beam_splitter}
\mathcal{H} = \omega(a_X^\dagger a_X + a_Y^\dagger a_Y)  + i\sqrt{2\gamma} (a_X^\dagger a_Y - a_Y^\dagger a_X). 
\end{equation}
We introduce quadratures $q = (a_X + a_X^\dagger)/\sqrt{2}$ and $p = i (a_X^\dagger - a_X)/\sqrt{2}$ (and similarly for $Q$ and $P$ for the ancilla). 
The Hamiltonian then becomes of the form~(\ref{H}), with $H_X = H_Y = \omega {\mathbb I}_2$ and 
\begin{equation}\label{beam_splitter_C}
C = \begin{pmatrix}
0 & -\sqrt{2\gamma} \\[0.2cm] \sqrt{2\gamma} & 0 \end{pmatrix} 
\end{equation}
As a consequence, the matrices $C_X$ and $C_Y$ in Eq.~\eqref{CXCY} become
\begin{equation}\label{beam_splitter_CSCA}
C_X = -C_Y =  \sqrt{2\gamma} I_2,
\end{equation}
so that  $A$ in~\eqref{gaussian_AAndD} becomes 
\begin{equation}\label{KS_example}
A = \omega \Omega_X - \gamma {\mathbb I}_2 = 
\begin{pmatrix}
-\gamma & \omega \\[0.2cm]
-\omega & -\gamma 
\end{pmatrix}. 
\end{equation}
The interaction with the ancilla therefore introduces a damping term of intensity $\gamma$. 

We also assume that the ancilla is initially thermal, 
\begin{equation}
\sigma_Y = (\bar{n}+\nicefrac{1}{2}) {\mathbb I}_2, 
\end{equation}
where $\bar{n}$ is the thermal occupation. 
The diffusion matrix $D$ then becomes
\begin{equation}
D = C_X \sigma_Y C_X\trans = \gamma(2\bar{n}+1) {\mathbb I}_2. 
\end{equation}
The resulting unconditional dynamics, compounding the effects of multiple collisions, therefore corresponds to the usual quantum optical master equation 
\[
\frac{d\rho_X}{dt} = -i [H_X,\rho_X] + \gamma(\bar{n}+1) \mathcal{D}[a_X] + \gamma \bar{n} \mathcal{D}[a_X^\dagger],
\]
where $\mathcal{D}[L] = L \rho_X L^\dagger - \frac{1}{2} \{L^\dagger L, \rho_X\}$.

\subsection{\label{app:example:QQ}Position-position coupling}

Next we consider the case in which the system and ancilla are still given by a single mode each, but now coupled through an interaction of the form 
\begin{equation}\label{position_position}
\mathcal{H}_\text{int} = - \sqrt{g} q Q.
\end{equation}
The interaction matrix $C$ in Eq.~(\ref{H}) becomes 
\begin{equation}\label{QQ_C}
C = -\sqrt{g} \begin{pmatrix} 1 & 0 \\[0.2cm] 0 & 0 \end{pmatrix},
\end{equation}
so that 
\begin{equation}\label{QQ_CXCY}
C_X = C_Y = \sqrt{g}\begin{pmatrix} 0 & 0 \\ 1 & 0 \end{pmatrix}.
\end{equation}
Quite interestingly we see that in this case $C_X C_Y = C_Y C_X = 0$. Hence, the drift terms in Eqs.~(\ref{gaussian_AAndD}) and (\ref{gaussian_AY}) vanish completely.
The diffusion matrix, on the other hand, becomes 
\[
D = g(\bar{n}+\nicefrac{1}{2}) \begin{pmatrix} 0 & 0 \\ 0 & 1 \end{pmatrix}.
\]
A position-position coupling therefore introduces diffusion only in the momentum quadrature (and no damping in either). 

\subsection{Ancillae with multiple components}

As a final example, let us consider the case where each interaction actually involves an ancilla with two components, $Y = (Q_1, P_1, Q_2, P_2)$. 
This helps to gain  intuition about the sizes of the matrices. 
It is also important when only some of the ancillae are actually measured, which is an experimentally meaningful hypothesis: normally, the system will interact with many ancillae at once, but the experimenter may have access to only some of them. 

The interaction matrix $C$ in Eq.~(\ref{H}) now becomes rectangular:
\begin{equation}
C = \begin{pmatrix} C_1 & C_2 \end{pmatrix}, 
\end{equation}
with $C_1$ and $C_2$ being $2\times 2$ matrices. 
The matrices $C_X$ and $C_Y$, in turn, become
\begin{equation}
C_X = \begin{pmatrix} C_{X_1} & C_{X_2} \end{pmatrix}, 
\qquad 
C_Y = \begin{pmatrix} C_{Y_1} \\[0.2cm] C_{Y_2}\end{pmatrix}, 
\end{equation}
where $C_{X_i}$ and $C_{Y_i}$ are all $2\times 2$. 
For instance, if $(Q_1,P_1)$ interacts with the system according to a beam-splitter interaction~\eqref{beam_splitter}, then $C_1$ will be given exactly by Eq.~\eqref{beam_splitter_C} and $C_{X_1}$, $C_{Y_1}$ will be given by Eq.~\eqref{beam_splitter_CSCA}.

%
%
\section{\label{app:gaussian_stuff} Calculation of information and thermodynamic quantities in the continuous variable scenario}
%
%

This appendix provides details on the calculation of information-theoretic and thermodynamic quantities in the case of continuous variable models. 
To do so we will use some of the results of Appendix~\ref{app:gaussian}. 
The information rate appearing in Eq.~\eqref{Delta_I_splitting} 
is given by 
\begin{equation}\label{gaussian_delta_I}
 \Delta I_t = \frac{1}{2} \ln \frac{ |\sigma_{X_t}| }{ |\sigma_{X_{t-1}} | } -
   \frac{1}{2} \ln \frac{ |\sigma_{X_t| \zeta_t}| }{ |\sigma_{X_{t-1}| \zeta_{t-1}}| } .
\end{equation} 
But to compute $G_t$ and $L_t$ in Eq.~\eqref{Delta_I_splitting} we need $I(X_t\! : \! \zeta_{t-1})$. 
This is obtained from the map
\begin{equation}\label{intermediate_unconditional_state}
\rho_{X_t|\zeta_{t-1}} = \mathcal{E}(\rho_{X_{t-1}|\zeta_{t-1}}), 
\end{equation}
where $\mathcal{E}$ is the unconditional map $\rho_{X_{t}} = \mathcal{E}(\rho_{X_{t-1}}) := \tr_{Y_t} \left\{\rho_{X_{t}Y_{t}'}\right\}.$ In the language of covariance matrices, this consists in applying the unconditional evolution~\eqref{gaussian_sigmaXZ_2} to $\sigma_{X_{t-1}|\zeta_{t-1}}$; viz., 
\begin{equation}
\sigma_{X_t|\zeta_{t-1}} = \sigma_{X_{t-1} |\zeta_{t-1}} + \Big(A \sigma_{X_{t-1}|\zeta_{t-1}} + \sigma_{X_{t-1}|\zeta_{t-1}} A\trans + D  \Big) dt.
\end{equation}
We then immediately get, using the definitions~\eqref{gain} and~\eqref{loss}, 
\begin{IEEEeqnarray}{rCl}
\label{gaussian_G}
G_t &=& \frac{1}{2} \ln \frac{ |\sigma_{X_t| \zeta_t}| }{ | \sigma_{X_t| \zeta_{t-1}} | }, 
\\[0.2cm]
\label{gaussian_L}
L_t &=& \frac{1}{2} \ln \frac{ |\sigma_{X_{t-1}}| }{ |\sigma_{X_t}| } -  \frac{1}{2} \ln \frac{ |\sigma_{X_{t-1}| \zeta_{t-1}}| }{ | \sigma_{X_t| \zeta_{t-1}} | }. 
\end{IEEEeqnarray}
As a sanity check, subtracting $G_t - L_t$ clearly leads to~\eqref{gaussian_delta_I}.

Eqs.~\eqref{gaussian_delta_I},~\eqref{gaussian_G} and~\eqref{gaussian_L} do not assume infinitesimal collisions. 
To obtain a continuous time description, we expand the determinants to leading order in $dt$.
To do this, the following result turns out to be quite useful:
Consider the Wigner entropy~\eqref{gaussian_renyi2} and assume that $\sigma = \sigma_0 + \sigma_1$, where $\sigma_1$ is small. 
A series expansion of  $|\sigma|$ in powers of $\sigma_1$ then yields,
\begin{equation}\label{Gaussian_entropy_expansion}
\frac{1}{2}\ln |\sigma| = \frac{1}{2}\ln |\sigma_0| + \frac{1}{2} \tr\big( \sigma_0^{-1} \sigma_1\big) - \frac{1}{4} \tr\big(\sigma_0^{-1} \sigma_1 \sigma_0^{-1} \sigma_1\big)+\ldots
\end{equation}
This is  a useful expression because all results just presented are of this form.

We begin by applying it  to Eq.~\eqref{gaussian_delta_I}.
First, from~\eqref{gaussian_uncond_sigma_discrete} we get 
\[
\frac{1}{2} \ln \frac{ |\sigma_{X_t}| }{ |\sigma_{X_{t-1}} | } = \frac{1}{2} \tr\big(2A + \sigma_{X_{t-1}}^{-1}~D\big) dt,
\]
where we used the fact that $\tr(A) = \tr(A^{\text{T}})$. 
Similarly, Eq.~\eqref{gaussian_sigmaXZ_t} yields
\[
\frac{1}{2} \ln \frac{ |\sigma_{X_t|\zeta_t}| }{ |\sigma_{X_{t-1}|\zeta_{t-1}} | } = \frac{1}{2} \tr\Bigg\{2A + \sigma_{X_{t-1}|\zeta_{t-1}}^{-1}~D~-~\sigma_{X_{t-1}|\zeta_{t-1}}^{-1}~\chi[\sigma_{X_{t-1}|\zeta_{t-1}}]  \Bigg\}dt. 
\]
In the limit $dt \to 0$, Eq.~\eqref{gaussian_delta_I} therefore reduces to the result in Eq.~\eqref{gaussian_dIdt}.
Similarly, repeating the  procedure for $G_t$ and $L_t$ in Eqs.~\eqref{gaussian_G} and~\eqref{gaussian_L} and identifying  $\dot{G} = G_t/dt$ and $\dot{L} = L_t/dt$, leads to Eqs.~\eqref{gaussian_G2} and~\eqref{gaussian_L2}. 

To compute the thermodynamic quantities, we first note that the  relative Wigner entropy between two Gaussian states $\rho_1$ and $\rho_2$, with covariance matrices $\sigma_1$ and $\sigma_2$, and first moments $r_1$ and $r_2$, can be written as ~\cite{Adesso2012}
\begin{IEEEeqnarray}{rCl}
D(\rho_1 || \rho_2) &=& \frac{1}{2} \tr\Big[ \sigma_2^{-1} (\sigma_1 - \sigma_2) \Big] + S(\sigma_2) - S(\sigma_1) \\[0.2cm]
\nonumber
&&+ \frac{1}{2} (r_1 - r_2)\trans \sigma_2^{-1} (r_1 - r_2).
\end{IEEEeqnarray} 
Thus, the entropy flux in a single collision can be written as 
\begin{equation}\label{gaussian_phi_0}
\Delta \Phi_t = \frac{1}{2} \tr\Big[ \sigma_{Y_t}^{-1} \big(\sigma_{Y_t'}- \sigma_{Y_t}\big) \Big] 
+\frac{1}{2} r_{Y_t'}\trans~\sigma_{Y_t}^{-1}~r_{Y_t'}. 
\end{equation}
Plugging in Eqs.~\eqref{gaussian_rY} and~\eqref{gaussian_sigmaY} then leads to Eq.~\eqref{gaussian_phi},
where $\dot{\Phi} = \Delta \Phi_t/dt$. We also need to use the fact that 
\begin{equation}\label{gaussian_A_trace}
\tr(A_Y) = \tr(A) = \frac{1}{2} \tr(C_X C_Y), 
\end{equation}
which follows from Eqs.~\eqref{gaussian_AAndD} and~\eqref{gaussian_AY}, together with the fact that $\Omega_X H_X$ and $\Omega_Y H_Y$ are traceless. 
Finally, to obtain $\dot{\Sigma}^u$, we use again the expansion~\eqref{Gaussian_entropy_expansion} to 
\begin{equation}\label{gaussian_unconditional_entropy_change}
S(X_t) = S(X_{t-1}) + \frac{1}{2} \tr \Big\{2A +  \sigma_{X_{t-1}}^{-1}D \Big\} dt.
\end{equation}
Combining this with~\eqref{gaussian_phi_0} and taking the limit $dt\to 0$ then leads to  Eq.~\eqref{gaussian_uncond_sigma}.

We finish with a technical note. 
In deriving these expressions we have tacitly assumed that $\Delta\Phi_t^c=\Delta\Phi_t^u$ is satisfied. 
In general, however, there is no guarantee that the state of the ancilla after the (non-selective) measurement, $\tilde{\rho}_{Y'}=\sum_z M_z\rho_{Y'}M_z $, will still be Gaussian. This is due to the fact that, despite the POVM of a (noiseless) general-dyne measurement corresponds to the PVM over some (pure) Gaussian state, there are infinitely many (unitarily equivalent) quantum operations corresponding to the same POVM and some of these operations can give rise to a non-Gaussian $\tilde{\rho}_{Y'}$. It is however easy to see that, any time the state of the ancilla $\tilde{\rho}_{Y'}$ is Gaussian, then the entropy flux is well-defined in terms of the Wigner relative entropy and $\Delta\Phi_t^c=\Delta\Phi_t^u$ is  automatically satisfied. Physically speaking, this  is always the case when we assume the operations acting on the ancilla to be the projections over Gaussian states. Moreover, since the state of the ancilla after the measurement is rarely of interest, and in accordance with the classical intuition spelled out  in~\cite{firstpaper}
, the possible mismatch between the conditional and unconditional fluxes can be safely neglected.

\bibliography{library}
\end{document}